\documentclass{amsart}
\usepackage{amsmath,amssymb,amsfonts,amsthm}
\usepackage{setspace} 
\usepackage{framed}
\usepackage{amssymb}
\usepackage{pgf}
\usepackage{amsmath}
\usepackage{amssymb}
\usepackage{verbatim}
\usepackage{listings}
\usepackage{sidecap}
\usepackage{float}
\usepackage{todonotes}
\usepackage{epsfig}
\usepackage{mathrsfs}
\usepackage{verbatim}
\usepackage[titletoc]{appendix}
\newtheorem{theorem}{Theorem}[section]

\newtheorem{proposition}[theorem]{Proposition}

\theoremstyle{definition}

\newtheorem{example}[theorem]{Example}

\theoremstyle{remark}
\newtheorem{remark}[theorem]{Remark}

\usepackage{lipsum}

\newcommand\blfootnote[1]{%
  \begingroup
  \renewcommand\thefootnote{}\footnote{#1}%
  \addtocounter{footnote}{-1}%
  \endgroup
}

\numberwithin{equation}{section}

\newcommand{\R}{\mathbb{R}}
\newcommand{\N}{\mathbb{N}}

\newcommand{\K}{\mathcal{K}}
\newcommand{\Q}{\mathbb{Q}}

\newcommand{\E}{\mathbb{E}}
\renewcommand{\P}{\mathbb{P}}
\newcommand{\PP}{\mathbb{P}}
\newcommand{\Qq}{\mathcal{Q}}

\newcommand{\QQ}{\mathbb{Q}}
\renewcommand{\d}{\ \mathrm{d}}

\usepackage{lipsum}

\begin{document}

\title[Matching distributions]{Matching distributions:\\ Derivatives pricing with physical density\\ shape correction}

\author{Jarno Talponen}
\blfootnote{University of Eastern Finland, Department of Physics and Mathematics, Box 111, FI-80101 Joensuu, Finland,
talponen@iki.fi}

\keywords{state price density, static hedging, derivative, pricing kernel, non-Gaussian, fat tails, skew, non-structural, implied distribution, distribution matching, payoff distribution pricing model\\
JEL classification: G10, G12, G13, C02}
\date{\today}

\begin{abstract}

In this paper a clean, simple and economically cogent computational method is introduced for correcting for excess kurtosis and skew in the pricing of European-style options. In fact, virtually any deviation in the physical distribution (e.g. from the Black-Scholes-Merton model) can be accommodated in a flexible, non-structural and semi-parametric fashion. 
The method does not involve expansions. It is based on a kind of statistical-static hedging technique related to Dybvig's (1988) distribution pricing. This gives rise to a state price density estimate $\widehat{q}$ with some tangible benefits. Its principle is transparent, and it is easy to implement numerically, while avoiding some typical issues involved in such an estimation. We will analyze the properties of this estimator and provide some justification for it. At the end we illustrate numerically how the Black-Scholes-Merton model can be flexibly accommodated with non-Gaussian physical distributions.
\end{abstract}

\maketitle


\section{Introduction}

\noindent 
In incomplete markets the prices of a new asset and derivatives on it may not be attainable by using proper hedging strategies. Even then statistical hedging may be applicable in reducing the asset's risk component with unknown risk premium. 
This may readily provide a reasonable model for the price of the asset by 
applying a simple correction, for instance Ross's APT pricing on the residual unhedgeable risk component.

The pricing of equities, on one hand, and derivatives, on the other hand, are usually treated as separate problems with rather different techniques. However, the estimation or calibration of the state price density (SPD) of a new asset, studied here, incorporates the price information of both the asset and the European style derivatives on it.

The function formed by taking state-by-state the ratio of the SPD and the physical density is known as the \emph{pricing kernel} or \emph{stochastic discount factor} (SDF). The pricing kernel is a central frequently applied tool in finance, and its unexpected empirical shape is studied by Bakshi et al. (1997), (2010) and Song and Xiu (2016). The estimated pricing kernel may be U-shaped, which is not in line with its interpretation as the marginal utility at equilibrium.
This connection is discussed by Beiglb\"{o}ck et al. (2012) and Reichlin (2013). 
The physical and risk-neutral characteristics of the observed densities have been connected by Chernov and Ghysels (2000), Bakshi 
et al. (2003), Chalamandaris and Rompolis (2012) and Engle and Figlewski (2015). The dependence between these densities remains interesting both from the theoretical and practical point of view. It is also the focus of this paper.

Several authors have considered the problem of correcting for skewness and leptokurtic effects of the equity returns in the Black-Scholes-Merton (BSM) and other pricing models. Gram-Charlier expansions are studied by Corrado and Su (1997), (2007), Knight and Satchel (2000), Longstaff (1995), Madan and Milne (1994), and discussed in the monograph by Jondeau et al. (2007) dedicated to the topic. The technique introduced here addresses the same issue, and some further possible uses are discussed at the end of the paper. The rough idea of the constructions is the same: one builds a new asset on the old benchmark model in such a way that some specifications, e.g. the physical moments, are satisfied by the newly modeled asset, which in turn has its new `internal' SPD.

Here static hedging techniques and statistical hedging philosophy are combined to construct a hedgeable proxy derivative for the non-hedgeable new asset. The economic rationale in using such proxies is that if two assets appear superficially sufficiently similar, then the investors may not differentiate between them in pricing, even though their returns do not strictly coincide as cash flows, which is the case in hedging based on a proper arbitrage. 
For instance, statistical arbitrage activity in a market conceivably leads to a situtation described in the APT. 
Dybvig (1988) considers a \emph{payoff distribution 
pricing model} (PDPM) where the price of an asset depends on its single-step payoff distribution. This work was recently extended by Rieger (2011) and Beare (2011) where there is further discussion on the devolepments around the PDPM. Dybvig is mainly interested in finding an extreme price range for a given payoff distribution. Although
the starting point is somewhat similar in this paper, the approach is eventually quite the opposite, since here we investigate rather conservative monotone rearrangements of the state space which seem reasonable in 'correcting' a benchmark model.

In this paper a SPD estimation technique is introduced for an asset highly correlated with a liquid proxy security, which in turn has a rich class of underlying European style options on it. 
A \emph{distribution matching pricing} principle is introduced here. It is easily described; \emph{one simply constructs by static hedging a European derivative on the proxy security such that the payoff distribution matches the price distribution of the asset being priced.} Thus, the asset is valued based on its price distribution at a given strike time of the derivatives. The constructed derivative and the security become \emph{comonotone}.
Ideally, the asset price $A_T$ becomes almost perfectly correlated with the proxy derivative payoff $f(S_T )$. 
Here $S_T$ is the proxy security price at the given strike $T$. 

This problem is mathematically ill-posed since for a given price distribution there exist several matching derivatives, martingales and prices. This serious issue is in part alleviated if the proxy derivative can be constructed in such a way that it is indeed \emph{highly correlated} with the asset. Thus there are fewer hypothetical martingales on the proxy security to consider. The new pricing model is not merely an arbitrary one satisfying the given physical density specifications, it is also in some sense statistically close to the benchmark model. Despite the theoretical obstructions, the correction for distributions in analytical European style option pricing models remains of practical importance.

The treatment here is continuous-state, rather than being discrete, such as in implied trees studied by Rubinstein (1994) and Monte Carlo methods reviewed in Glasserman (2003) and Jaeckel (2002). Instead, the approach taken here follows to some extent the general philosophy of Bakshi et al. (2003) and Jarrow and Rudd (1982). 
The former raises the problem of differential pricing of individual equity options versus the market index. 
This can be addressed by using our main formula \eqref{eq: 1} for comparisons of SPDs. 
These matters are also closely related to the works of Buchen and Kelly (1996), Halperin and Itkin (2014), Hocquard et al. (2015) and Madan (2006). 

The main benefits of the distribution matching are the following.
It is essentially model-free, semi-analytic form and does not require any \emph{ad hoc} discretizations or families of density functions. In particular, it does not require assumptions on the dynamics of the asset, which often play a major role in econometrics,
see e.g. Andersen et al. (2015). This approach is stable even under infinite variance, a case which occurs typically in connection with fat-tailed distributions. The technique here can be seen as a non-structural, semi-parametric (cf. Stutzer 1996) version of \emph{moment matching} techniques investigated by Airoldi (2005) and Brigo et al. (2004). Instead of matching some first moments of the physical distributions, the distributions are \emph{completely} matched. In particular, the technique accommodates all skewness-kurtosis pairs, unlike Rubinstein's 
(1998) Edgeworth trees and Johnson binomial trees investigated by Simonato (2011). It admits even multi-modal risk-neutral and physical distributions. Also, the issue of negative probabilities does not arise here.

The formula obtained for the new asset's SPD is simple:
\begin{equation}\label{eq: 1}
\widehat{q}_1 (x) = \frac{\phi_{1}(x)}{\phi_{2}(\K(x))} q_2 (\K(x)). 
\end{equation}
where $\phi_2$ and $q_2$ are the benchmark model physical density and SPD, respectively, and $\phi_1$ is the new model's  physical density. Model $1$ may be considered as the 'corrected' version of pricing model $2$. 
The state space transform $\K$ is increasing, it preserves physical probabilities between the models and can be 
easily computed numerically.

The above SPD appears as a useful by-product from a proxy derivative construction. As an intermediate step, this proxy derivative is first approximately assembled using only finite portfolios of digital options on the proxy security. This shows how statistical-static hedging of a given asset could be accomplished in practice.The resulting asset pricing rule by such proxy derivatives is not linear in general, cf. Chateauneuf et al. (1996), but it is homogenous (respects the scaling of cash flows), respects a Modigliani-Miller type separation and satisfies some other natural properties, like continuity and monotonicity with respect to stochastic dominance of assets.

The inter-model formula \eqref{eq: 1}, which connects physical densities with state prices in a simple way, is the crux of this investigation. It may be interesting in connection with the volatility smile, pricing kernel shapes, non-parametric calibration of SPDs and related matters. Although \eqref{eq: 1} should be considered an estimate in general, it performs correctly for risk-neutral densities if the pricing models are sufficiently isomorphic. For instance, this is the case if we compare BSM models having the same market price of risk. See Section \ref{sect: soundness} where we provide some facts which support \eqref{eq: 1}.

Interestingly, it turns out that the state space transform $\mathcal{K}$, arising purely from the proxy derivative construction, in fact induces an optimal transport plan for the given physical measures. This is a central notion in the theory of optimal transport, which in turn has been applied recently in various economic problems. These include martingale optimal transport, economic equilibration, \emph{matching} problems and hedonic models, see e.g. 
Henry-Labordere (2017), Ghoussoub and Moameni (2014) for discussion. Hence the name of this paper.

As an application, the inter-model formula is applied in correcting for kurtosis and skew deviations in the BSM model. The numerical algorithm is easy to implement and is illustrated at the end. The considerations also appear to lead to some further interesting econometric analysis, which could be termed as 'Implied Physical Distributions', generalizing implied volatility.

\section{Preparations}

The references Bakshi et al. (2003) and Jarrow and Rudd (1982) provide the context for this paper.
For a suitable background information, see A\"{\i}t-Sahalia and Lo (1998),
Breeden and Litzenberger (1978), Carr and Chou (1997), Cochrane (2005), Derman et al. (1995), Fengler (2005), 
F\"ollmer and Schied (2005), Jondeau et al. (2007) and Shreve (2004). 

\subsection{Preliminaries on the formalism}

First a word of warning: we often discuss matching a cash flow. Here we seldom hedge cash flows by proper arbitrage. Instead, we typically build a matching cash flow in \emph{distribution}, which is a considerably weaker hedging notion, 
cf. Dybvig (1988). 
The terminology 'estimate' and the corresponding notation are used rather liberally for the sake of convenience. In particular, the errors are not specified statistically and no unbiasedness etc. are claimed. 

We will consider mostly a single-step model with the present time $t$ and the future maturity of options $T$. 
As usual, the physical measure involving asset prices is denoted by $\PP$. We will denote by $\Qq$ the state price measure and by $\QQ$ the risk-neutral measure, both assumed absolutely continuous with respect to the Lebesgue measure on the state space. For instance, the price of an asset and bond are 
\[S(t) = \int x \d\Qq = \int x \d\Qq(S(T)=x),\quad  B(t) = \int 1 \d\Qq.\]
Recall that if the densities of measures $\mu$ and $\nu$ on the real line are continuous functions 
$\phi$ and $\psi$, respectively, then the Radon-Nikodym derivative is
\[\frac{\d \mu}{\d \nu} (x) =\frac{\phi (x)}{\psi (x)}.\]
For convenience we assume throughout all densities to be continuous and their supports to be (possibly unbounded) intervals.
In case of standard Lebesgue measure $m$ we surpress $\d x = \d m (x)$.
\newcommand{\q}{\mathbf{q}} 
The SPD and the risk-neutral density (RND) on an asset $S$ are denoted respectively by
\[q(x)= \frac{\d\Qq(S(T)=x)}{\d x},\quad  \q(x)= \frac{\d\Q(S(T)=x)}{\d x}.\]
As usual, the time-$t$ value of a 'plain vanilla' European call option $C$ on security $S$ is denoted by 
$C (S_t ,t,T,K)$. Similarly, we denote the price of a 'digital call' $C_{digi}$ having payoff $1_{S(T)\geq K}$.
Here $t\geq 0$ is time of valuation, $T$ is the expiration and $K$ is the strike price. More generally, $V$ stands for the value of portfolios and assets at a given time.
Let us recall the \emph{market price of risk}, $\frac{\mu-r}{\sigma}$,
where $\mu$, $r$ and $\sigma$ are the usual parameters of the BSM model.

\subsection{The valuation method explained}
We will first price an asset $S_1$, considered as a single-step stochastic cash flow, by constructing a derivative security with a matching payoff distribution. Thus the valuation involves statistical hedging in comparing the distributions and static hedging in constructing a suitable derivative in the single-step framework. However, by no means is such a derivative unique, nor is the price unique. Therefore, we are required to make some further specifications.

Pricing by distribution matching involves an asset being valued, $S_1$, and a benchmark security, $S_2$ which is traded
and has abundantly sorts of European style options written on it. The technique is focused on a given time interval $[0,T]$ and in particular the distributions of $S_1 (T)$ and $S_2 (T)$ which involve separate models, $\mathcal{M}_1$,  
$\mathcal{M}_2$. More precisely, these single-step models are 
\[\mathcal{M}_i = (\Omega, \phi_i , q_i),\quad i=1,2\]
where the state space $\Omega$ of $S_1 (T)$ and $S_2 (T)$ is considered $\R_+$ here 
and $\phi_i$ and $q_i$ are the physical and state price densities on $\Omega$. The time $T$ is also the expiry of the options on $S_2$. 

Ideally, the asset being valued, i.e. $S_1$, and the benchmark security $S_2$ are very highly correlated and with approximately same return distributions, that is, the asset $S_1$ is a 'quasi twin security' of $S_2$. In a less ideal situation we have non-perfectly, but still highly correlated liquid proxy security \emph{and} European style call options on the proxy security. We will use the option price information here to patch some of the information lost due to imperfect correlation.

This is performed as follows. The SPD of the proxy security can be estimated from option prices. This density can be interpreted as a system of 'infinitesimal' Arrow-Debreu securities, or degenerate double digital options. Intuitively, we will reweigh the A-D securities and reassemble them to obtain a portfolio with the same distribution as the asset being priced. 
The portfolio of A-D securites, which will be highly correlated with the asset, can be viewed as the derivative sought after.
   
To outline the distribution matching pricing method, we consider $S_1$, an asset to be valued, and a proxy security $S_2$. We take as given time-$t$ estimates of the following: 
\begin{enumerate}
\item The continuous physical distributions $\phi_1$ and $\phi_2$ on $S_1 (T)$ and $S_2 (T)$, 
\item The SPD $q_2$ on $S_2 (T)$. 
\end{enumerate}
The distribution matching technique then yields $\widehat{q}^{\mathrm{DM}}_{S_1}$, 
an estimate for the SPD on $S_1 (T)$.    
   
This technique boils down to SPD transformations. The particular transforming, or the A-D securities reassembly procedure, is performed in such a way that the resulting prices meet some natural 'rationality' conditions. 

The SPD $q_2$ may or may not follow some standard option pricing formula, and it may be estimated by a specialist from the market data. See e.g. the works Ait-Sahalia and Lo (1998) and Ait-Sahalia and Duarte (2003).

In fact, this approach can be viewed as an asset pricing counterpart of real options valuation, where cash flows are modeled by trees which can be solved. However, the analysis here is continuous-state and 
there is active calibration according to market information.

Next we will explin the assumptions, or rather the thought experiment behind distribution matching. This is also related to implied trees and Marketed Asset Disclaimer in real options analysis.

In distribution matching one first considers portolios $\Pi_0$ of digital options on $S_2$ with expiry $T$ in such a way that the value of the portolio $\Pi_0(T)$, considered at time $t=0$, is highly correlated with $S_2 (T)$ and the distribution of the portfolio value at time $T$ is close to that of $S_1 (T)$, in symbols 
\[\mathrm{Corr} (\Pi_0 (T), S_2 (T))\approx 1, \quad \Pi_0 (T) \sim S_1 (T)\ \text{approximately}.\]
These portfolios are finite and as such the distributions are rough approximations of that of $S_1 (T)$, and
at this stage the portfolios are by no means unique.

As a response to the above issues one passes to the limit in the process of refining these portfolios in such a way that the payoff distribution of the portfolio matches \emph{exactly} the distribution of $S_1 (T)$. 
This leads to an analysis of physical and state price distributions with a particular state space transformation.
The resulting idealized portfolio $\Pi$ of infinitesimal Arrow-Debreu securities is not \emph{ad hoc} anymore at this stage; instead it is a unique arrangement of options such that some natural properties of the pricing functional are satisfied.

Namely, the portfolio $\Pi$ can be seen as a European style derivative on $S_2$ such the payoffs $S_2 (T)$ and $\Pi (T)$ are comonotone. 

In an ideal case where the benchmark security and the asset are very highly correlated, the ideal portfolio $\Pi$ then 
satisfies
\begin{equation}\label{eq: fund}
\mathrm{Corr} (\Pi (T), S_1 (T))\approx 1, \quad \Pi (T) \sim S_1 (T).
\end{equation}

Note that $\Pi (T)$ then statistically hedges $S_1$ but we argue the hedging notion is actually much stronger due to high 
correlation. If the correlation was in fact perfect, then the portfolio would be a complete hedge for $S_1$ in the single-step model. 

The thought experiment is reasonable if we consider the known ill-posed problem of correcting the BSM model for deviations from lognormality. In such an analysis it appears a natural starting point to consider markets which accommodate two highly correlated and approximately similarly distributed assets: One which exactly follows the BSM model and another one with only approximately normal return rates.

The technical details are discussed next. It would be conventional to consider mainly risk-neutral measures in the analysis, but we adopt a different, albeit completely equivalent, approach, considering mainly ideal portfolios of Arrow-Debreu securities. We choose to do so, since the latter approach seems more transparent in this setting.

\subsection{Continuous portfolios of infinitesimal Arrow-Debreu securities}\label{sect: SPD}

Let us recall some well-known ideas of Breeden and Litzenberger (1978).
See also the related works of Bick (1982), Brown and Ross (1991) and Jarrow (1986).
Consider the price $C_{digi}(S(t),t,T,K) $ of digital call options with payoff $1_{S_T \geq K}$.
Suppose that $\frac{\partial^2 C_S}{\partial^2 K}$ exists and is continuous on $K$ (although the Breeden-Litzenberger representation generalizes to a more general setting). Then, regardless of the model, this represents the SPD:
\[q_S (K)=\frac{\partial^2 C(S(t),t,T,K)}{\partial^2 K} (K) =-\frac{\partial C_{digi}(S(t),t,T,K)}{\partial K}(K) .\]

Next we will consider a formal digital option on $S$, which pays $1$ unit of numeraire if the underlying asset satisfies
$K_1 \leq S(T) \leq K_2$, and pays $0$ in the contrary case.  
We will apply a formal notation
\begin{equation}\label{eq: AD_dR}
AD(S(t),t,T,\d K) = q_S (K)\d K
\end{equation}
which becomes sensible in the context of integration.
The above corresponds intuitively to the degenerate case $K_1=K= K_2$ where the trigger is exactly $S(T)=K$ or 
$K_2-K_1 =\d K$ is negligible. Such an option is subsequently termed an \emph{infinitesimal Arrow-Debreu security}.
The chances of this option being triggered are negligible as well, so the value of it is 'infinitesimal', hence the terminology and the right hand term $\d K$.

We wish to form portfolios consisting of infinitesimal Arrow-Debreu securities with all possible strikes at the same time.
Thus these portfolios are not only infinite, but they contain continuum many types of assets. 
The information on the distribution of different types of calls can be conveniently decoded as a Radon measure. We will formalize these portfolios as signed absolutely continuous measures, denoted by $\rho$. Let us denote the value of the portfolio by $\Pi(t) =\Pi_\rho (t)$.

If the Radon-Nikodym derivative $\frac{\d \rho}{\d K}(K)$ is positive at a point $K$ this means that 
the portfolio has a long position on the Arrow-Debreu security corresponding to strike $K$. 
Similarly, if $\frac{\d \rho}{\d K}(K)$ is negative, then the portfolio has a short position on the A-D security corresponding to strike $K$, and, moreover, the relative weight of the position is $|\frac{\d \rho}{\d K}(K)|$. It is instructive to think of relative weights as being analogous to probability densities, only the sign may vary according to the short/long position. 

The payoff of such a portfolio $\rho$ at time $T$ in the case $S(T)=K$ is
\[[\Pi (T)\ | \ S(T)=K]\ =\ \frac{\d \rho}{\d K}(K).\]
The financial interpretation here is that the portfolio is a bundle of A-D securities with different strikes, and all other securities, except the ones with strike $K$, expire worthless. Thus the payoff of the portfolio is the amount of strike-$K$ A-D securities held. 

The value of the portfolio at time $t<T$ is the aggregate value of all the A-D securities in it:
\[\Pi (t) = \int \frac{\d \rho}{\d K}(K)\ AD(S(t),t,T,\d K).\]
The financial interpretation is that $AD(S(t),t,T,\d K)$ is the price of $1$ strike-$K$ A-D security and $\frac{\d \rho}{\d K}(K)$ is the amount of such securities in the portfolio.

Thus, if we wish to construct a European style derivative with payoff $f$ (which could also have negative values, e.g. in case of futures contracts), we assemble a portfolio $\rho$ with the weights $\frac{\d \rho}{\d K}(K) = f(K)$ for each strike $K$.
The value of the porfolio then assumes a familiar form:
\begin{multline*}
\Pi (t) = \int \frac{\d \rho}{\d K}(K)\ AD(S(t),t,T,\d K) = \int f(K)\ q(K) \d K \\
= \int q(K) f(K) \d K = \int \frac{AD(S(t),t,T,\d K)}{\d K}(K)\  \d \rho(K) .
\end{multline*}
For instance, a standard European call option with strike price $K_0$ is replicated by a portfolio $\rho$ as follows: 
For each $K>K_0$ there are included $K-K_0$ many strike-$K$ A-D securities, thus
$\frac{\d \rho}{\d K}(K) =(K-K_0 ) 1_{K>K_0} $.

This idea discussed above is surely not new to specialists, but this tool will be useful in performing static replication transparently.

\renewcommand{\P}{\mathbb{P}}
\renewcommand{\Q}{\mathbb{Q}}

\section{The pricing framework}
Let us consider two assets, an asset that will be priced, $S_1$, and a proxy asset $S_2$. We will study $2$ different 
pricing models $\mathcal{M}_i$ corresponding to these assets. When comparing the SPDs of the models we usually 
assume the short rates in the models coincide, $r_1=r_2$, so that the SPDs $q_1$ and $q_2$ integrate to the same bond price. 

Suppose that $\phi_{1},\phi_{2}\colon \R\to [0,\infty)$ are continuous density distributions of $S_1 (T)$ and $S_2 (T)$. 
We assume that the supports are intervals $[a_i ,\infty)$, $i=1,2$. We denote by $\P_1$ and $\P_2$ the corresponding physical probability measures\footnote{It is debatable whether the assets, modeled as random variables, should coexist in a same probability space 
or not. If they do, then these measures can be seen as push-forward measures. This issue is analogous to the \emph{marketed asset disclaimer} in real options analysis.}.  

\renewcommand{\K}{\mathcal{K}}

Thus there is an absolutely continuous increasing function $\K\colon (a_1 ,\infty)\to (a_2,\infty)$ such that $\P_2 (\K(I))=\P_1 (I)$ (where 
$\K(I)$ is the image of $I$) for any interval $I \subset (a_1 ,\infty)$ \footnote{Hence for any measurable subset by a Dynkin-type argument.}. The above probability-preserving condition can be stated for an increasing and absolutely continuous map $\mathcal{K}$ equivalently as 
\[\P_{1} (S_1 (T) \leq K) =  \P_{2} (S_2 (T) \leq \K(K))\ \ \text{for each}\ \ K> a_1 .\]

The purpose of $\K$ is to pair up the distributions in a suitable way. This is certainly not the only possible 
$\P_1$-$\P_2$-measure-preserving transformation (see Dybvig 1988), but it appears to be a natural one which leads to reasonable conclusions.

To replicate the value distribution of $S_1$ we will construct a suitably weighted portfolio $\rho$ of 
Arrow-Debreu securities of varying strikes on the state space of $S_2$. 
Our aim is to build a portfolio $\rho$ whose time-$T$ value, given the event $S_2 (T)=\K(S_1 (T) )$ for any $S_1 (T) >0$, satisfies
\[[\Pi (T)\ |\ S_2 (T)=\K(S_1 )]\ =\ \frac{\d\rho}{\d K}(\K(S_1 ))=S_1 (T). \]
In other words, the portfolio $\rho$ 'contains $S_1 (T)$ many' A-D securities corresponding to 
the event $S_2 (T)=\K (S_1 (T))$ where the corresponding A-D securities are essentially 
degenerate double digital options on $S_2$ with payoff $1_{S_2 (T)=\K(S_1 (T))}$. Thus the portfolio of 
A-D securities pays exactly $S_1 (T)$ in the event $S_2 (T)=\K(S_1 (T))$. Since the transformation 
$\K$ is order-preserving and $\P_1$-$\P_2$-measure-preserving, it follows that for every $a<b$ we have
\[\P_{1,t} (a< S_1 (T) < b) = \P_{2,t} (a< \Pi_\rho (T) <b ).\]
Thus the value distribution functions of $S_1 (T)$ and $\Pi (T)$ coincide.

\subsection{Intermediate stage: Approximating the cash flow with finite portfolios of digital options}\label{sect: fin_approx}
To make the analysis more tangible and cogent we begin the construction of $\K$ with 
an intermediate step involving simple portfolios. This provides the means required to statistically hedge assets 
by static replication strategy containing long and short positions of (finitely many) potentially traded options on the proxy security. Eventually we will pass on to the limit, letting the number of steps $n$ used in discretization tend to infinity and thus asymptotically match the distribution of $S_1$. Nachman (1988) studies the approximation of general European derivatives in a static model with portfolios of plain vanilla options.

For convenience, let us assume at this stage that the supports of $\phi_{i}$ are bounded intervals. Then we may approximate in distribution the value $S_1 (T)$ with a finite portfolio of digital calls. Let $\varepsilon>0$. Then there is an $n\in \N$ with the following partitions: 
\begin{enumerate}
\item $x_{1}<x_{2}<\ldots <x_{n}$ of the support of $\phi_{1}$ such that $\int_{x_k}^{x_{k+1}}\phi_{1}\ \d x=1/n$ and $x_{k+1}-x_{k}<\varepsilon$ for $k=1,\ldots,n-1$.
\item $y_{1}<y_{2}<\ldots < y_{n}$ of the support of $\phi_{2}$ such that 
$\int_{y_k}^{y_{k+1}}\phi_2\ \d x=1/n$ for $k=1,\ldots,n-1$.
\end{enumerate}
Clearly 
\[x - \sum_{k=1}^{n-1} \frac{x_{k}+x_{k+1}}{2}1_{[x_{k},x_{k+1}]}(x)\]
is bounded by $\varepsilon$, uniformly for all states $x=S_1 (T)$.
Motivated by this observation, we will build a portfolio on the security side to match the above linear combination 
of indicator functions. We form a portfolio of double digital calls by including in for each $k=1,\ldots,n-1$ a position with payoff
\begin{equation}\label{eq: aver}
\frac{x_{k}+x_{k+1}}{2}1_{[y_{k},y_{k+1}]}
\end{equation}
which has the replication cost
\begin{equation}
\frac{x_{k}+x_{k+1}}{2}(C_{digi}(S(t),t,T,y_k)-C_{digi}(S(t),t,T,y_{k+1})).
\end{equation}
The corresponding weights on the digital options are $\frac{x_{1}+x_{2}}{2}$ 
and $\frac{x_{n-1}+x_{n}}{2}$ for strikes $y_1$ and $y_n$, respectively, and  
\begin{equation}
-\frac{x_{k}+x_{k+1}}{2}+\frac{x_{k+1}+x_{k+2}}{2}=\frac{x_{k+2}-x_{k}}{2} 
\end{equation}
for other strikes $y_{k+1}$. Here we may define the corresponding portfolio of A-D securities $\rho$ by 
\[\rho(A) = \sum_{k=1}^{n-1}\frac{x_{k}+x_{k+1}}{2}\int_{y_k }^{y_{k+1}} 1_{A}(y)\d y.\]
Indeed, note that the arbitrage-free value of the instrument
\[C_{digi}(S(t),t,T,y_k)-C_{digi}(S(t),t,T,y_{k+1})\] 
is $\int_{y_k}^{y_{k+1}} 1\d\Qq$.
The price of the described portfolio at time $t$ is 
\begin{equation}\label{eq: double_star}
\Pi (t) =\sum_{k=1}^{n-1}\frac{x_{k+1}+x_{k}}{2}\int_{y_k}^{y_{k+1}} 1\d\Qq.
\end{equation}

By cultivating the above portfolio construction with discretized version of \eqref{eq: dR} below, one may build 
an approximating portfolio with plain calls as well.

\section{Constructing a cash-flow-distribution-equivalent portfolio}
Recall that the density of the event $S_1 (T)=x$ is $\phi_{1}(x)$.
The corresponding event on the proxy security side is $\K(x)$, and its density is $\phi_{2}(\K(x))$. In the case of such 
an event the corresponding A-D security returns $1$ unit of numeraire (and not $x$ units). Therefore we will compensate by using the weight $x$ on the A-D security corresponding to the event $\K(x)$. Secondly, 
\[\K(x+\Delta x)-\mathcal{K}(x)+o(\Delta x) = \Delta x \frac{\phi_{1}(x)}{\phi_{2}(\K(x))}=\Delta x \frac{\d\P_1}{\d K}(x)/\frac{\d\P_2}{\d K}(\K(x)),\]
since $\K$ is measure-preserving. Here the fractions on the right hand side are the Radon-Nikodym derivatives of the measures. Thus, (see \eqref{eq: AD_dR}) we will regard $\K$ as a weak solution to the differential equation
\begin{equation}\label{eq: dR}
\K'(x)=\frac{\phi_{1}(x)}{\phi_{2}(\K(x))},\ \K(a_1 )=a_2,\ x>a_1 .
\end{equation}
In practice, the densities are continuous and the solution is in the usual sense. Note that $\K$ can be computed easily by 
numerically solving the above separable ODE.
If $F_{1}$ and $F_2$ are the corresponding cumulative distributions, then
\begin{equation}\label{eq: transp} 
\K(x)=F_{2}^{-1}(F_{1}(x)).
\end{equation} 
There is an interesting digression independent of the subsequent financial motivations. Equation
\eqref{eq: transp} defines a well-known solution\footnote{The above arrangement minimizes so-called Wasserstein's $W_p$ distances on the real line, see e.g. Svetlozar and Ruschendorf (1998).} to a problem in mathematical optimal transport theory, motivated by logistics and mathematical economics. Such problems were first considered rigorously
by Kantorovich (1942).

For each $x>a_1$ the cash flow $x1_{[x,x+\Delta x]}$ (considered a contingent claim on $S_1$) can be 'matched' via transform
$\K\colon \R\to \R$, up to precision $o(\Delta x)$, by a portfolio (a contingent claim on $S_2$) with value 
\begin{equation}\label{eq: diffe}
x\ C_{digi}(S(T),T,T,\K(x))-x\ C_{digi}(S(T),T,T,\K(x+\Delta x)).
\end{equation}
Thus, in both the cases the possible payoffs are the same ($0$ and $x$), and the probabilities of the 
positive outcomes coincide. 

\newcommand{\DM}{\mathrm{DM}}

In our portfolio we will buy at time $t$ the amount $x$ of (infinitesimal) A-D securities at each strike 
$\K(x)$: 
Then $\rho$ can be characterized by 
\[\frac{\d \rho}{\d K}(\K(x))=x.\]
Recalling the chain rule of differentiation, \eqref{eq: diffe} and the Breeden-Litzenberger representation considerations, 
we obtain the time-$t$ value of portfolio $\rho$ of A-D options. 
\begin{proposition}
In the above setup the arbitrage-free time-$t$ price of $\rho$ is
\begin{equation}\label{eq: value}
\begin{array}{lll}
\Pi_\rho (t)&=&\int_{a_1 }^\infty x  \left( -\frac{\partial C_{digi}(S_2 (t),t,T,K)}{\partial K}\Big\vert_{K=\K(x)}\right) \frac{\d\K}{\d x}\ \d x\\
& &\\
&=&\int_{a_1 }^\infty x\ q_2 (\K(x))\ \frac{\phi_{1}(x)}{\phi_{2}(\K(x))} \d x.
\end{array}
\end{equation}
\end{proposition}

Here the portolio can be interpreted as a European style derivative on $S_2$ with the following properties:
\begin{enumerate}
\item At time $t$ the derivative at maturity $T$ and $S_1 (T)$ have the same value distribution. 
\item The payoff of the derivative is an absolutely continuous strictly increasing function on the value $S_2 (T)$. 
\end{enumerate} 
The latter condition typically implies that the derivative payoff and $S_2 (T)$ are highly correlated.  

Provided that all the relevant information is understood, we denote by $\widehat{S}^{\DM}_1 (t)$ the value \eqref{eq: value}. To summarize (see subsequent Proposition \ref{prop: contclaim}):
\begin{proposition}\label{prop: simple} 
Assume the above setup. If $S_1$ is a European style derivative on $S_2$ with absolutely continuous strictly increasing payoff, 
$S_1 (T)=f(S_2 (T))$, then the arbitrage-free price of $S_1$ is
\[S_1 (t)= \widehat{S}^{\DM}_1 (t) .\]
\end{proposition}
Thus, if $S_1$ can be regarded as being approximately such a derivative, then pricing $S_1$ boils down to valuing the above derivative 
and an 'error term asset', possibly by some other means, e.g. the APT. 

Above we applied for simplicity the same discount factor in both pricing models. If we are comparing two models with 
a priori known or assumed discount factors, which are different, then the Arrow-Debreu assets must be additionally scaled, so that, for instance, the risk-free bond
prices in asset-$1$ model become correctly priced. This leads to 
considering risk-neutral densities (RND) $\q_i$ in place of SPDs $q_i$.

Then \eqref{eq: value} yields a natural estimate for the unobserved RND on $S_1$:
\begin{equation}\label{eq: estQ}
\widehat{\q}^{\mathrm{DM}}_1 (x)=\frac{\d \widehat{\Q}_1}{\d x} (S_1 (T)=x) = \frac{\phi_{1}(x)}{\phi_{2}(\K(x))} \q_2 (\K(x)).
\end{equation}
By rearranging we may interpret the above in a more familiar way:
\[\frac{\d\widehat{\Q}_1}{\d\P_1}(x) = \frac{\d\Q_2}{\d\P_2}(\mathcal{K}(x))\]
which states that the pricing kernels of the pricing models coincide, up to a transformation of states, and, rearranging and by  \eqref{eq: dR}, we have
\begin{equation}\label{eq: rearrnage}
\frac{\widehat{\q}^{\mathrm{DM}}_1 (x) }{\q_{2} (\mathcal{K}(x))}=\frac{\phi_{1} (x)}{\phi_{2} (\mathcal{K}(x))}= \mathcal{K}'(x) .
\end{equation}


\section{Distribution matching asset valuation: Basic properties}\label{sect: fund}
There are potentially several possible measure-preserving transformations to choose from. However, the transformation $\K$, which continuously preserves the order of states, appears heuristically the most reasonable. The claim that this transform is 'natural' is also corroborated by the following nice features which are specific to this particular type of transform.  

\begin{proposition}[Limit of approximate portfolios]\label{prop: properties}
Assume that $\phi_1$ and $\phi_2$ are as above.
Then the values of the approximating finite portfolios in \eqref{eq: double_star} converge to the value \eqref{eq: value}.
Moreover, a similar conclusion holds in case where the supports of $\phi_i$ are unbounded intervals if the averages 
(see \eqref{eq: aver}) are replaced by smaller absolute value terms of the respective subintervals.
\end{proposition}

The following fact is an immediate result of the construction of the portfolios in distribution matching.

\begin{proposition}[Uniqueness up to distribution]\label{prop: idemp}
Assume that the distributions $S_1 (T)$ and $S_2 (T)$ coincide, $\phi_{1}=\phi_{2}$, and the proxy security $S_2$ has a SPD in its model. As above, suppose that we are using the information of $S_2$ model with the distribution of $S_1 (T)$ to price $S_1$ by distribution matching. Then 
\[\widehat{S}^{\DM}_1 (t) = S_2 (t) .\] 

\end{proposition}

\begin{proposition}[Monotonicity w.r.t. stochastic dominance]\label{prop: monotonicity}
Suppose that $S_a$ and $S_b$ are securities such that $S_a (T) \preceq S_b (T)$ 
and we are using the same proxy security $S_2$ model for both of them separately in distribution matching. 
Then 
\[ \widehat{S}^{\DM}_a (t) \leq \widehat{S}^{\DM}_b (t) .\]
\end{proposition}

Along the same lines one can show that if  $S_a (T) \prec S_b (T)$, 
then $\widehat{S}^{\DM}_a (t)  <  \widehat{S}^{\DM}_b (t) $.
We note that the fact that the state space transformation $\K$ is increasing is essential here. The pricing rule does not, however, preserve second order stochastic dominance without further assumptions.

\begin{proposition}[Continuity]\label{prop: cont}
Assume that the SPD $q_{2}$ corresponding to $S_2$ asset is bounded and $\frac{q_2 (y)}{\phi_2 (y)}\to 0$ as $y\to\infty$.
Suppose that assets $S_{1, (n)} (T) \to S_1 (T)$ in $\P_1$-mean as $n\to \infty$. Suppose that we apply distribution matching technique with $q_2$ and $\phi_1 , \phi_{1, (n)}$, $\phi_2$ in forming the corresponding portfolios $\rho$, $\rho_n$, respectively where $\widehat{S}^{\DM}_1 (t) <\infty$. Then 
\[\widehat{S}^{\DM}_{1, (n)} (t)  \overset{n\to\infty}{\longrightarrow} \widehat{S}^{\DM}_1 (t) .\]
\end{proposition}

The pricing method is \emph{not} linear, that is, if $S_a$ and $S_b$ are securities with respective 
A-D portfolios $\rho_{a}$ and $\rho_{b}$, then the portfolio $\rho$ resulting from the combined cash flow
$S_{a}+S_b$ typically satisfies $\rho\neq \rho_a + \rho_b$ and typically 
$\widehat{S_{a}+S_b}^\DM \neq \widehat{S_a}^\DM + \widehat{S_b}^\DM$. This is due to the fact that the valuation 
machinery, on the asset (to be priced) side, depends only on the distribution of the cash flow and does not take into 
account correlations.
Indeed, consider for instance equally distributed flows $S_a , S_b$ and $S_c$ such that 
$S_a + S_b$ has zero variance and $S_b + S_c$ has non-zero variance.
However, a Modigliani-Miller type separation of value holds, see Proposition \ref{prop: MM} below. 

\begin{proposition}\label{prop: contclaim}
Suppose that asset $S_1$ is in fact a European style contingent claim on asset $S_2$. We consider a model of $S_2$ 
with given $\phi_2$ and $q_2$. Assume further that the payoff $f_{S_1} (S_2 (T) )$ is absolutely continuous and strictly increasing on $S_2 (T)$. Then the distribution matching method prices $S_1$ correctly; the arbitrage-free price $S_1 (t)$ coincides with the value $\widehat{S}^{\DM}_1 (t)$. 
\end{proposition}

This has the following rather immediate consequences. Pricing by distribution matching is in a sense consistent within the BSM framework. 
\begin{remark}[Consistency with the BSM model]\label{prop: }
Consider a BSM model asset $S_2$ with a European option payoff $f$ and an asset $S_1$. Assume that $f$ and $S_1$ are as in Proposition \ref{prop: contclaim}. Then $\widehat{S}^{\DM}_1 (t)$ coincides with the initial value of the 
BSM value process of the derivative replication
\[ \widehat{S}^{\DM}_1 (t) = V_f (t) . \]
\end{remark}
The suitable payoff function $f$ appears in the construction of the required portfolio $\rho$ above, namely,
$f=\K^{-1}$. 

\begin{proposition}\label{prop: MM}
Suppose that at time $t$ the future prices $S_0 (T)$ and $S_1 (T)$ are perfectly correlated with the prices $S_2 (T)$. This means that these are obtained from each other by affine transforms (i.e. shifted linear transforms) and $S_0$ and $S_1$ can be viewed as European style options with strictly increasing dependence on the underlying asset $S_2$. 
Then distribution matching correctly prices $S_0$ and $S_1$, considered derivatives on $S_2$ with strike time $T$ and
the pricing is linear in this case:
\[\widehat{(a S_0 + b S_1 + c B)}^\DM (t) =  a\widehat{S}^{\DM}_0 (t) +   b \widehat{S}^{\DM}_1 (t) + cB(t),\quad a,b\geq 0.\]
\end{proposition}
The reason we require positive weights $a$ and $b$ and perfect correlation is that the distribution matching 
technique does not take into account the correlation structure of assets.

\section{Applications: Performing a correction in the pricing of European calls under skew and fat tails}

The above valuation technique suggests a method for correcting for any type of distribution deviation in a given analytic-form pricing model. The pricing of derivatives based on the price distribution of the underlying is plausible according to 
Proposition \ref{prop: simple}. However, without any information on the dynamics of the uderlying, the problem is ill-posed since there may be several martingale measures and corresponding arbitrage free prices. We assume that in addition
to the analyzed underlying asset $A$ in the market there is a highly correlated `quasi-twin' asset $A^*$ which accurately follows the dynamics of the analytic-form pricing model. Then the distribution matching technique produces a derivative
with payoff $f(A_{T}^*)$ which is equally distributed and highly correlated with $A_T$. Here $A$ is statistically hedgeable but may not be properly hedgeable by using derivatives on $A_{T}^*$. Thus there is a hypothetical 
`error term' cash flow $A^{\varepsilon}$ such that 
\begin{equation}\label{eq: A_T}
A_T = f(A_{T}^*) + A^{\varepsilon}_T
\end{equation}
where $A^{\varepsilon}_T$ has mean $0$ and small variance. In modeling it seems reasonable 
to treat the value $A^{\varepsilon}_0$ asymptotically as $0$, e.g. by invoking the CAPM, APT or other factor models, 
possibly non-linear ones, see Atlan et al. (2007).

The benefit of this method, similarly as in non-structural techniques mentioned in the introduction, is that we may amalgamate the desirable properties of empirically realistic models of asset returns, and, on the other hand, those of analytically tractable models. 

For example, we may attach skew and fat tail features in pricing European stock options, using the BSM model as the 'ground model'. To accomplish this we require all the data appearing in the BSM model at time $t=0$, 
excluding $\mu$ and $\sigma$, and a modeled physical (non LogNormal) distribution of the stock price at $T$. In the case of European type call we thus require the following data:
\begin{enumerate}
\item Asset price at time $t=0$, $S(0)$. (Given empirically.)
\item Interest rate $r$ at time $t=0$. (Empirically observed; we choose it similarly as if applying it in the BSM model.)
\item Maturity $T$. (Given.)
\item Strike price $K$. (Given.)
\item Modeled probability density function for the price of the stock at $t=T$, $S(T)$. (Either analytical form fitted to data or directly from data by local regression.) 
\end{enumerate}

{\it Setting the implied volatility and trend.} The implied volatility $\sigma$ and the implied trend $\mu$ are chosen in such a way that the median of the modeled physical distribution of $S(T)$ coincides with the median of the log normal physical distribution of the BSM model with parameters $S(0)$, $T$, $r$ and $\sigma$. Additionally, we require that the interquartile ranges (IQR) (i.e. the lengths of the intervals) coincide. The median and the IQR clearly determine $\sigma$ and $\mu$ uniquely.

The motivation for doing this is twofold. Firstly, considering medians appears compatible with the way we constructed the derivative above. That is, by transforming distributions in states' order-preserving and continuous fashion, so that there is one-to-one correspondence between the quantiles of the probability distribution and its transformed version. Secondly, suppose that we have a PDF of the form $f=\alpha g + (1-\alpha)h$ where $0< \alpha < 1$, $\alpha \approx 1$, $h$ is log normally distributed and $h$ is also a very skewed and fat tailed one. Thus $f$ is a kind of mildly modified version of $g$ with some added skew and fat tails. This case corresponds to a typical application here. Note that the median and IQR of $f$ are close to that of $g$. Thus the internal calibration of $\sigma$ and $\mu$ by equating the medians and IQRs. Also note that using the means or variances of the distributions would be out of the question, since the mean of the modeled physical distribution may fail to exist, due to fat tails, but, on the other hand, the quantiles of a distribution always exist.

\subsection{Illustration: Parametric case. Modeling the risk-neutral distribution from a given mixed LogNormal-LogCauchy-LogStudent-L\'evy  type physical distributions superposed on the BSM model}

To illustrate the application of the pricing scheme under investigation we computed the prices of calls under varying underlying physical distributions with negative skew and excess kurtosis.

We approximate numerically the SPDs $q$ in a relatively dense grid. We denote by $\phi_{Mix}$ the mixed 
PDF and by $\phi_{BSM}$ (resp. $q_{BSM}$) the log normal physical PDF (SPDs) of the BSM model. We compute the corresponding SPDs $\widehat{q}_{Mix} = q$ by the following loop in pseudo code:\\

\vbox{
\begin{framed}
{\bf
Let $x_1 = h$ and fix $y_1$ such that $\int_{0}^{y_1} \phi_{BSM} (x)\ dx= \int_{0}^h \phi_{Mix}(x)\ dx$;\\

For i=1 to N;\\

\quad Let $q(x_{i})=q_{BSM}(y_{i}) \frac{\phi_{Mix}(x_i )}{\phi_{BSM} (y_i )}$;\\

\quad Let $x_{i+1}=x_i +h$ and $y_{i+1}=y_{i}+ h \frac{\phi_{Mix}(x_i )}{\phi_{BSM} (y_i )}$;\\

Next;\\

}
\end{framed}
}

To torture our model with super-heavy tails, we consider physical distributions which are mixtures of the following distributions:  LogNormal, LogCauchy, LogStudent ($\nu=2$), L\'evy. 
The main component is the lognormal distribution with BSM model parameters $t=0$, $T=1$, $S(0)=1$, $r=0.04$,  
$\mu_0 =0.1$, $\sigma_0 =0.3$ and the parameters of the rest of the distributions are in the program documentation. 
We formed the mixed distribution by taking a weighted average of the distributions, putting the same weight on all the non-LogNormal distributions and the rest of the weight on the LogNormal distribution. 
We performed the calculations under different weights $w=0.00,\ 0.01,\ 0.02,\ 0.05,\ 0.10$ involving the fat-tailed distributions. Thus the respective LogNormal weights were $1.00,\ 0.97,\ 0.94,\ 0.85,\ 0.70$. 
Note that the LogCauchy density is unbounded which affects the mixed distributions as well. This puts some burden on the numerical implementation.

We identified the median and the quartile interrange (QIR) of each of the mixed distributions corresponding to different weights. We then searched for LogNormal distribution parameters, $\mu$ and $\sigma$, such that the median and the QIR coincide for the mixed distribution and the LogNormal one.
We use the same values for $t,T,r$ and $S(0)$ fixed previously. We are not assuming here that $S(0)$ is the correct price of the underlying in question, it is merely the price of a 'nearby' proxy security, 
possibly a hypothetical one. The BSM model fixed above is then used as a ground model which provides the required densities $\phi_{BSM}$ and $q_{BSM}$.

For the weight $w=0$ and strike price $K=0$ we get the value $S(0)=1$ for the call option, as one expects. Changing the assumptions regarding the physical distribution, which in this framework is reflected by the risk neutral distributions, affects the modeled prices of the underlying asset (according to \eqref{eq: estQ}) against the proxy security. 

The formation of the modeled state-price density is illustrated below. On the $x$-axis we have the states of the underlying asset $S_1 (T)$ where the benchmark security (corresponding to the case $w=0$) is at-the-money at $1$. For numerical reasons we report states $x\geq 0.2$ since $\mathcal{K}'$ has a singularity at $x=0$.

\begin{figure}[h!]
\centering
\includegraphics[width=\textwidth]{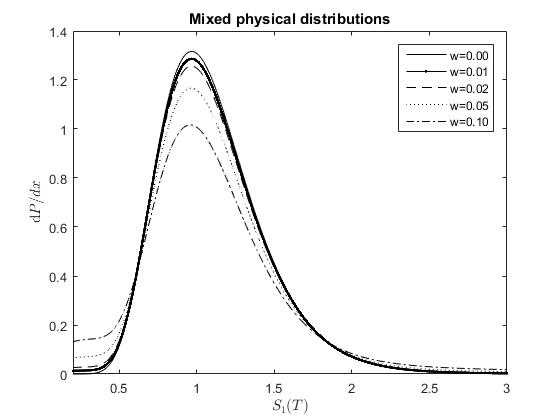}
\caption{The kurtosis increases along with the relative weight of the fat-tailed distributions in mixed distributions. The modes of the distributions are 
$\approx 0.97$ (excluding the left tails).}
\end{figure}

\begin{figure}[h!]
\centering
\includegraphics[width=\textwidth]{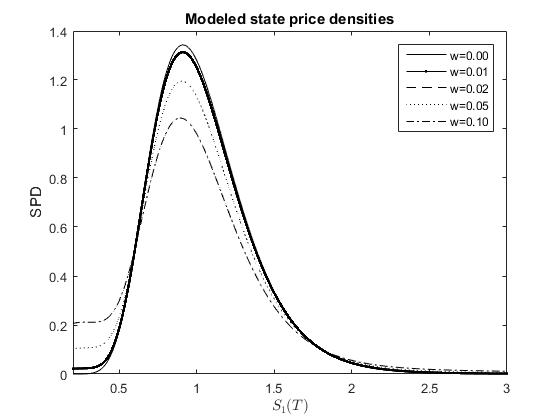}
\caption{The resulting SPD estimates $\widehat{q}_{Mix}^{\mathrm{DM}}$ resemble the physical distributions. Note that the state prices are rather large at left hand tail.
These can be interpreted as risk-neutral densities $\d\mathbb{Q}/\d x$, up to normalizations by the bond price. The modes of  the risk-neutral distributions are $\approx 0.91$ (excluding the left tails).}
\end{figure}

\begin{figure}[h!]
\centering
\includegraphics[width=\textwidth]{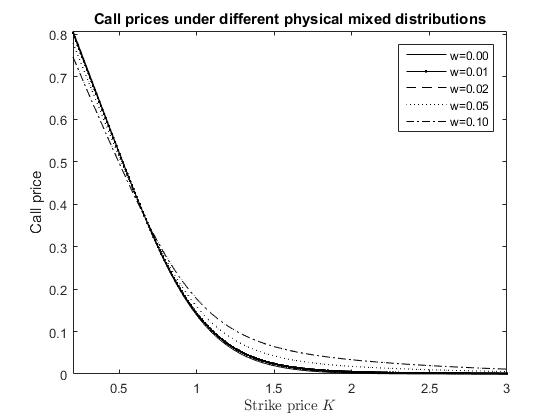}
\caption{The heavy left hand tail decreases the price of the underlying (the $K=0$ case) and the heavy right hand tails increase the price of calls with 
high strike price.}
\end{figure}

\begin{figure}[H]
\centering
\includegraphics[width=\textwidth]{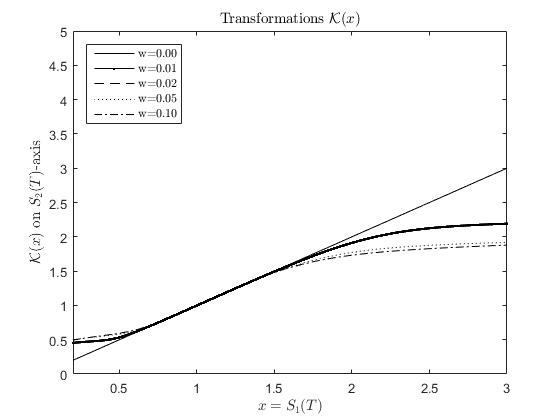}
\caption{The transformations $\mathcal{K}(x)$ (apart from the identity, the $w=0$ case) appear discontinuous at $0$ which is not the case. The right hand values tend to infinity very slowly. Both these features are due to heavy tails. The approximate coincidence of the medians is due to the setup of the distributions.}
\end{figure}

\begin{figure}[H]
\centering
\includegraphics[width=\textwidth]{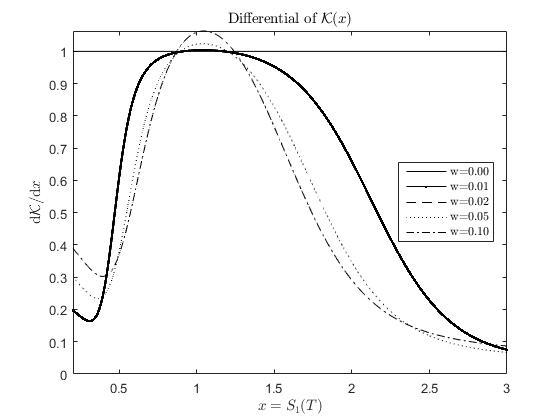}
\caption{The differential $\mathcal{K}' (x)$ blows up at $0$. The small values on the right hand side reflect the fact that the ratio $\phi_{Mix}(x)/\phi_{BSM}(\mathcal{K}(x)) \equiv 1$ is enforced while $\phi_{Mix}$ is heavy-tailed and $\phi_{BSM}$ is not.}
\end{figure}

\begin{figure}[H]
\centering
\includegraphics[width=\textwidth]{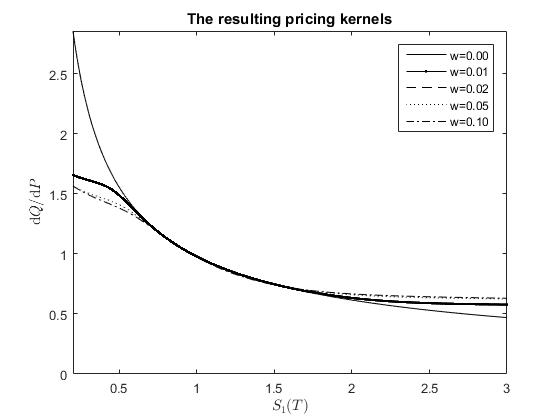}
\caption{Distribution matching with heavy tails has a tendency to spread the states around at-the-money states 
with moderate risk-neutral/physical densities ratio. This also illustrates that distribution matching does not reduce to an application of pricing kernel in conjunction with the heavy tailed empirical density.}
\end{figure}

The differential $\mathcal{K}' (x)$ has a singularity at $0$. For the sake of numerical stability of the algorithm, the interval $[0,0.01]$ had to be treated separately and the plot does not include this interval. This does not cause an economically significant error, since the probabilities and the prices of the stock are small on this interval anyway.

The computations were performed on Matlab 2015b (8.6.0.267246 win64). The accuracy of numerical integration and Euler's method for solving ordinary differential equation performed was heuristically controlled by varying the step size $h=10^{-k}$. The computations required about $1$ minute on a laptop computer.

In the above illustration we took the initial benchmark security price as given, $S_2 (0)=1$. Then we priced all European 
style options on $S_1$, including the trivial case with the asset itself, $S_1 (0)$. However, if the price of the asset 
$S_1 (0)$ is observed, then we may 'calibrate' $S_2 (0)$ so that the method produces the right observed price 
$S_1 (0)$. Indeed, recall that in the construction of the pricing measure the asset $S_1$ is modeled as a European derivative on security $S_2$ with increasing payoff function. Thus for instance in the BSM model the price $S_1 (0)$ is a strictly increasing continuous function of the initial price $S_2 (0)$.

\section{The soundness of the inter-model formula}\label{sect: soundness}
Eventually we will apply our technique in correcting for the skew and kurtosis in BSM pricing model. This problem is in fact ill-posed, the way it is considered here, since we are not making any assumptions on the dynamics of the asset prices. Admittedly, this fact may bring into question the soundness of the valuation technique. 

We argue that formula \eqref{eq: 1} provides a reasonable estimate for model $1$ SPD. To justify this we will sketch some situations where it performs accurately.

The assumptions made below are eventually rather restrictive. However, this does not mean that the pricing technique, understood as a reasonable estimate, should be equally restricted in its applicability. 

Distribution matching also resembles some valuation methods in capital budgeting. 
Namely, the \emph{marketed asset disclaimer} of real options analysis (see Trigeorgis 1999) is analogous to the proxying principle applied here.

\subsection{Comparison to risk-neutral pricing}
Let us assume that we have an asset modeled as a single-step cash flow
\[A_T = f(A_{T}^*) + A^{\varepsilon}_T \]
as in \eqref{eq: A_T}. If $\mathrm{Var}_t (A_T)<\infty$ then we may use 
$f(A_{T}^*) = \E_t (A_T \ |\ A_{T}^* )$ almost surely and then 
\[A^{\varepsilon}_T = A_T -  \E_t (A_T \ |\ A_{T}^* )\]
becomes uncorrelated with $A_{T}^*$, $ \E_t (A^{\varepsilon}_T ) =0$ and $\mathrm{Var}_t (A^{\varepsilon}_T ) <\infty$. Then, in the setting of Proposition \ref{prop: contclaim}, we may price $A$ up to the error term,
\[\E^{\Q_2}_t (A_T )= \E^{\Q_2}_t ( f(A_{T}^*)) + \E^{\Q_2}_t (A^{\varepsilon}_T )\]
if $A$ is a European-style derivative on $A^*$ in model $2$. 
The first risk-neutral expectation on the right-hand side can be correctly priced by Distribution Matching
\emph{if} the above payoff $f$ is assumed to be absolutely continuous and increasing. The second expectation, or the pricing error, can be controlled if model $2$ is specified, so that $\E_t ((\frac{d\Q_2}{d\P_2})^2 )$ can be calculated,
\[
\left|\E^{\Q_2}_t (A^{\varepsilon}_T )\right| \leq \E^{\Q_2}_t (|A^{\varepsilon}_T | ) 
=\E_t \left(\frac{d\Q_2}{d\P_2} |A^{\varepsilon}_T |\right)
\leq \sqrt{\E_t \left(\left(\frac{d\Q_2}{d\P_2}\right)^2 \right)\ \E_t \left( |A^{\varepsilon}_T |^2 \right)}. 
\]
Here we applied the Cauchy-Schwarz inequality and $\sqrt{\E_t \left( |A^{\varepsilon}_T |^2 \right)}$ is the 
standard deviation of $A^{\varepsilon}_T$.

\subsection{Digression: Value-at-Risk threshold utility correspondence between market models with representative agents}

In this section we study a special case where the SPDs of the market models are determined 
by utility functions of representative investors. 
The conditions for aggregating the markets by means of the notion of a representative investor have been studied e.g. 
by Ait-Sahalia and Lo (2000) and Rubinstein (1974). Recall that in pricing with market equilibrium and representative agent utility function a first order optimality condition involving the equilibrium can be expressed 
as follows:
\[\\q_{S_T} (K)  = c u' (K) \phi_{S_T} (K),\]
where the left hand side denotes the SPD of the state $S_T =K$, $u$ is the utility function of the 
representative agent and $c>0$ is a suitable constant depending on the time value of money and the marginal utility of the initial capital at time $t$. 

Suppose that $u_a$ and $u_b$ are differentiable representative agent utility functions of market models 
$\bf a $ and $\bf b $, respectively. Usually the single-attribute utility functions are thought to depend on the absolute level of money (or numeraire, or consumption level), but alternatively they may be written in the form 
\[u(x)=\nu (\Phi (x))\]
where $\nu$ is a suitable differentiable strictly increasing auxiliary function and $\Phi$ is the cumulative distribution of $S_T$. This can be arranged as a convention, since there is typically a $1$-to-$1$ correspondence between the states $K$ of $S_T$ and the cumulative probabilities. In terms of Value-at-Risk
essentially the same can be expressed by stating that the map $\alpha \mapsto \mathrm{VaR}_\alpha$ 
(confidence level $\alpha$ to the respective quantile), together with the given utility function induce the 
auxiliary function $\nu(\alpha)=u( -\mathrm{VaR}_\alpha)=u(\Phi^{-1} (\alpha))$. Writing the utility function in this form is merely a matter of convention.

Suppose that the representative agents partially agree on their utility in the sense that there is a universal way the utility functions are formed, only the market model specific physical distributions differ:
\begin{equation}\label{eq: uaub}
\begin{array}{ll}
u_a (x)=c_a \nu (\Phi_a (x)),\quad q_a \Big|_{S_{T}^{(a)} =x} =c_a \frac{\d}{\d x} \nu (\Phi_a (x)),\\
u_b (x)=c_b \nu (\Phi_b (x)), \quad q_b \Big|_{S_{T}^{(b)} =x } = c_b \frac{\d}{\d x} \nu (\Phi_b (x)).
\end{array}
\end{equation} 
This means that both the utility functions are values of a common auxiliary value profile $\nu$ parametrized according to confidence levels 
and 
\[c_b u_a (\mathrm{VaR}^{(a)}_\alpha ) =c_a u_b (\mathrm{VaR}^{(b)}_\alpha ).\]

This is one instance where the distribution matching, pairing the market models $\bf a $ and $\bf b $, can be applied to correctly predict the prices in one of the models with respect to that of the other. This essentially follows from the chain rule:
\[q_a \Big|_{S_{T}^{(a)} =x} = c_a \frac{\d}{\d x} \nu (\Phi_a (x)) = c_a \nu' (\Phi_a (x)) \phi_a (x), \]
\[q_b \Big|_{S_{T}^{(b)} =x} = c_b \frac{\d}{\d x} \nu (\Phi_b (x)) = c_b \nu' (\Phi_b (x)) \phi_b (x) .\]

Note that if the values of maturity $T$ face value $1$ risk-free zero-coupon bonds at time $t=0$ coincide in the 
market models $\bf a $ and $\bf b $, then
\[B_a (t)=\int c_a \nu' (\Phi_a (x)) \d\Phi_a =  \int c_b \nu' (\Phi_b (x)) \d\Phi_b =B_b (t), \]
so that $c_a=c_b$. Even if this were not the case, we may consider risk-neutral probability measures $\QQ_a$ and $\QQ_b$, in place of the pricing measures. Thus, we may essentially suppress the constants, $c_a =c_b=1$.

Now, if $\K$ is taken to be increasing, continuous and $\P_a$-$\P_b$-measure-preserving on the positive real line, we have that $\Phi_a (x) = \Phi_b (\K(x))$. Thus 
\[\q_b \Big|_{S_{T}^{(b)} = \mathcal{K}(x)} =  \frac{\d}{\d s} \nu (\Phi_b (s))\Big|_{s=\mathcal{K}(x)}
=  \nu' (\Phi_a (x)) \phi_b (\mathcal{K}(x))\]
so that
\[\q_a (S_{T}^{(a)} =x) =\frac{\phi_a (x)}{\phi_b (\mathcal{K}(x))}\q_b (S_{T}^{(b)} =\mathcal{K}(x) ).\]
This condition clearly states that \eqref{eq: value} correctly prices the $\bf a $ model asset 
in terms of the European call price system in $\bf b$ model, up to discounting. More generally, in this case the price of any European style option written on the $\bf a $ asset can be recovered from the $\bf b$ market model together with the physical distribution of the $\bf a $ asset.
\bigskip

\begin{example}\label{example}
Let us compare two separate BSM models with the same market 
price of risk:
\begin{equation}\label{eq: lambda}
\frac{\mu_a - r_a}{\sigma_a}=\frac{\mu_b - r_b}{\sigma_b} =\lambda.
\end{equation}
The required transform $\mathcal{K} \colon S_{T}^{(a)} \mapsto S_{T}^{(b)}$ is defined by
\[\frac{\log(S_{T}^{(a)}) - \log(S_{0}^{(a)}) - (\mu_a -\sigma_{a}^2 /2 )T}{\sigma_a} 
= \frac{\log(S_{T}^{(b)}) - \log(S_{0}^{(b)}) - (\mu_b - \sigma_{b}^2 /2 )T}{\sigma_b}\] 
after surpressing some insignificant terms. Indeed, this is clearly increasing, continous and it is also measure
preserving by the basic properties of the normal distribution. 

Note that 
\[\log(S_{T}^{(b)}) = \frac{\sigma_b}{\sigma_a} \log(S_{T}^{(a)}) + d\]
for a suitable constant $d$, since all the BSM model parameters and $S_{0}^{(a)}$, $S_{0}^{(b)}$ are deterministic. This means that
\begin{equation}\label{eq: K}
S_{T}^{(b)} = \mathcal{K}(S_{T}^{(a)}) = e^d (S_{T}^{(a)})^{\frac{\sigma_b}{\sigma_a}}.
\end{equation}
Note that direct calculations involving the risk-neutral and physical BSM model densities
yield 
\[\frac{\d\mathbb{Q}_i}{\d\mathbb{P}_i}(S^{(i)}_T) =e^{(r_i-\mu_i )\frac{\ln(S^{(i)}_T  /S^{(i)}_0 )}{\sigma_{i}^2}+\frac{(\mu_i -r_i )T}{2\sigma_{i}^2}},\quad i=a,b.\]
Next we will follow the approach where a utility function is implied by matching equilibrium first-order utility condition and the pricing kernel. Thus
\[\frac{\d\mathbb{Q}_a}{\d\mathbb{P}_a}(S^{(a)}_T) = u_{a}' (S^{(a)}_T),\quad 
\frac{\d\mathbb{Q}_b}{\d\mathbb{P}_b}(S^{(b)}_T) =  u_{b}' (S^{(b)}_T)\]
where we have isoelastic utility functions
\[u_a (S^{(a)}_T ) = c_1 (S^{(a)}_T )^{1-\frac{\mu_a -r_a}{\sigma_{a}^2}}
,\quad u_b (S^{(b)}_T ) = c_2 (S^{(b)}_T )^{1-\frac{\mu_b -r_b}{\sigma_{b}^2}},\]
so
\[\frac{\d\mathbb{Q}_a}{\d\mathbb{P}_a}(S^{(a)}_T ) = \frac{c_1}{1-\frac{\mu_a -r_a}{\sigma_{a}^2}} (S^{(a)}_T )^{-\frac{\mu_a -r_a}{\sigma_{a}^2}},\quad 
\frac{\d\mathbb{Q}_b}{\d\mathbb{P}_b}(S^{(b)}_T ) = \frac{c_2}{1-\frac{\mu_b -r_b}{\sigma_{b}^2}} (S^{(b)}_T )^{-\frac{\mu_b -r_b}{\sigma_{b}^2}}.\]
According to \eqref{eq: lambda} and \eqref{eq: K} we obtain
\begin{multline*}
(S^{(b)}_T )^{-\frac{\mu_b -r_b}{\sigma_{b}^2}}=(\mathcal{K}(S^{(a)}_T))^{-\frac{\mu_b -r_b}{\sigma_{b}^2}}
=((S_{T}^{(a)})^{\frac{\sigma_b}{\sigma_a}})^{-\frac{\mu_b -r_b}{\sigma_{b}^2}}
=(S_{T}^{(a)})^{-\frac{\lambda}{\sigma_{a}}}=(S^{(a)}_T)^{-\frac{\mu_a -r_a}{\sigma_{a}^2}},
\end{multline*}
up to multiplicative constants which we will continue surpressing bluntly. Recall that we are working in risk-neutral price system
which means that at the end the multiplicative factors are normalized suitably.
Put
\[\nu (\alpha) = (\Phi_{a}^{-1}(\alpha) )^{1-\frac{\mu_a -r_a}{\sigma_{a}^2}}.\]
Then 
\[u_{a} (x)= \nu (\Phi_{a} (x))=(\Phi_{a}^{-1}(\Phi_{a} (x)) )^{1-\frac{\mu_a -r_a}{\sigma_{a}^2}}
=x^{1-\frac{\mu_a -r_a}{\sigma_{a}^2}},\]
\[u_{a}' (x) = x^{-\frac{\mu_a -r_a}{\sigma_{a}^2}} =  x^{-\frac{\lambda}{\sigma_{a}}}\]
and 
\[u_{b}' (\mathcal{K}(x)) = x^{-\frac{\mu_b -r_b}{\sigma_{b}\sigma_a }}=x^{-\frac{\lambda}{\sigma_{a}}}=u_{a}' (x).\]
Then we obtain, given a European derivative payoff profile $f$, that  

\begin{multline*}
\E_{\Q_a}( f(S^{(a)}_T)) =\E_{\P_a}\left(\frac{\d\mathbb{Q}_a}{\d\mathbb{P}_a}(S^{(a)}_T )\  f(S^{(a)}_T)\right)
=\E_{\P_a}\left(u_{a}' (S^{(a)}_T)\  f(S^{(a)}_T)\right)\\ 
= \int u_{a}' (S^{(a)}_T)\  f(S^{(a)}_T) \frac{\d \P_a}{\d S^{(a)}_T}\d S^{(a)}_T 
= \int u_{b}' (\mathcal{K}(S^{(a)}_T))\  f(S^{(a)}_T )\  \frac{\d \P_a}{\d S^{(a)}_T}\d S^{(a)}_T\\
= \int  \frac{\d\mathbb{Q}_b}{\d\P_b}(\mathcal{K}(S^{(a)}_T))\  f(S^{(a)}_T )\  \frac{\d \P_a }{\d S^{(a)}_T}\d S^{(a)}_T
=\int \q_b (\mathcal{K}(x))\ f(x)\  \frac{\phi_a (x)}{\phi_b (\mathcal{K}(x))}\d x .
\end{multline*}
This means that in our setting with two separate BSM models with the same market price of risk 
the SPD estimation technique (see \eqref{eq: estQ}) performs correctly between the models for risk-neutral prices.
The short rates in the models $\bf a$ and $\bf b$ may of course differ and hence 
the discount terms must be reconciled. 
\end{example}

Next we will extend these ideas to the dynamical setting.

\subsection{Sufficiently isomorphic pricing models}\label{sect: paths}

Up to this point we have studied the behavior of assets essentially with a single time step and next will consider assets with continuous evolution of prices. We will study conditions relating the dynamics of the securities ($S_1$ and $S_2$ above) together, which guarantee that our pricing method indeed connectes the models together properly, see \eqref{eq: transf}.

The result in this section roughly states that if one has a \emph{misspecified} model for a security, then, under suitable conditions, one can still \emph{recover} the true value of the European style contingent claim on the security. 
To manage this we are clearly required to have some extra information and to assume some things on the model(s). Namely, assuming that the stochastic asset price models within the class are 'sufficiently isomorphic' we proceed by using distribution matching, thus coupling the \emph{biased} SPD and physical density together with the \emph{true} physical density. Then we recover the true SPD.

We analyze a case where two assets dynamically depend on a common latent stochastic state variable.
Suppose that $S_1 (t)$ and $S_2 (t)$ are driven by a common process $X_t$ with almost surely continuous realizations and
\[\frac{\d S_i}{S_i}(t) =  \mu_i (t) \d t + \sigma_i (t, S_i (t))\d X_t  .\]
We denote by $\mathcal{F}_t$ a filtration generated by the above stochastic processes and we assume it to be 
continuous. Here we assume that functions $\sigma_i >0$ and $\mu_i$ are continuous.

Suppose that the price processes can be approximated by binomial models $\mathcal{M}^{(n)}_i$ corresponding to $2^n$ even discrete increments $\vartriangle\!\! t = \frac{T}{2^n}$ in time  where the above stochastic differential equations are replaced by discrete type difference equations described shortly. We consider $2$-valued random variables $\Theta^{(n)}_{k\vartriangle t}$ (ups ($\Theta^+$) and downs ($\Theta^-$) in the binomial models)  such that 
$X^{(n)}_{t}  \to X_t$ in probability as $n\to\infty$ for every $t$, where $X^{(n)}_{t}$ is defined by piecewise linear 
interpolation from the values
\[X^{(n)}_{k\vartriangle t} = X_0 + \sum_{l \leq k}  \Theta^{(n)}_{l\vartriangle t}.\]

This means that $\{X^{(n)}_{k\vartriangle t}\}_k$, $n\in\N$, are binomial processes. Intuitively speaking,
$X^{(n)}_{t}$ are asymptotically adapted to $X_t$.
Define the corresponding assets 
\[S_{i}^{(n)} (0) =S_i (0),\]
\[S^{(n)}_{i, (k+1)\vartriangle t}-S^{(n)}_{i, k\vartriangle t} = S^{(n)}_{i, k\vartriangle t}\  \left(\mu_i (k\vartriangle\! t) \vartriangle\! t \ + \sigma_i (k\vartriangle t , S^{(n)}_{i, k\vartriangle t})\ \Theta^{(n)}_{(k+1)\vartriangle t}\right) .\]

In the literature it is customary to use scaling $\sqrt{\vartriangle\! t}$ in innovations in binomial models. However, we may resort to the above definition since the scale of the innovation terms $\Theta^{(n)}$ is not fixed here a priori. 
This provides us with the states and the corresponding probabilites
of the binomial models $\mathcal{M}^{(n)}_i$. Let us consider instantaneous short rates $r_i (t, S_i (t))$ and in the binomial
model $\mathcal{M}^{(n)}_i$ let the short rate used in the step $k\vartriangle\! t \leadsto (k+1)\vartriangle\! t$ be 
$r_i (k\vartriangle\! t , S^{(n)}_{i, k\vartriangle\! t})$.
 
The above seems to suggest approximating stock prices $\tilde{S}_{i}^{(n)}$ which are obtained by piecewise linear interpolation from the binomial processes:
\[\d \tilde{S}^{(n)}_i  (t) = S^{(n)}_{i, k\vartriangle t} \left(\mu_i (k\vartriangle\!\! t) +  \sigma_i (k\vartriangle t , S^{(n)}_{i, k\vartriangle t})\ \Theta^{(n)}_{(k+1)\vartriangle t}\big/ \vartriangle\! t\right)\d t ,\]
$k\vartriangle\! t <  t <(k+1)\vartriangle\! t$, so that 
\[\frac{\d \tilde{S}^{(n)}_i}{\tilde{S}^{(n)}_i}(t) \approx  \mu_i (t)\d t + \sigma_i (t , \tilde{S}^{(n)}_i )\ \d X^{(n)}_{t} ,\quad
k\vartriangle\! t <  t <(k+1)\vartriangle\! t \]
where $\d X^{(n)}_{t} = \Theta^{(n)}_{(k+1)\vartriangle t}\big/ \vartriangle\! t\ \d t$.

Instead, we define $S^{(n)}_i$ directly by 
\[S_{i}^{(n)} (0) =S_i (0),\]
\[\frac{\d S^{(n)}_i}{S^{(n)}_i}(t) =  \mu_i (t)\d t + \sigma_i (t, S^{(n)}_i )\ \d X^{(n)}_{t} ,\quad
k\vartriangle\! t <  t <(k+1)\vartriangle\! t \]
where the trajectories are absolutely continuous almost surely.
The dynamics appear very close to that of $S_i$ (notice the superscript $(n)$) but here the definition actually runs by means of classical analysis since the realizations of $X^{(n)}_{t}$ are piecewise linear.

\begin{theorem}\label{thm: coincide_MPR}
Let us consider the setting described above. We additionally assume the following stability conditions:
\begin{enumerate}
\item The physical and risk-neutral distributions of $S_{i} (T)$, $\phi_i$ and $\q_i =\frac{\d \QQ_i}{\d S_i (T)}$, are continuous functions on $(0,\infty)$.
\item $S_{i,T}^{(n)} \to S_i (T)$, $S^{(n)}_i (t)  \to S_i (t)$ in probability as $n\to\infty$ for all $t$. 
\item The unique risk-neutral probabilities of the states $S_{i,T}^{(n)}$ in models $\mathcal{M}^{(n)}_i$ converge in distribution to $\q_i$ in the sense that
\[\sum_{S_{i,T}^{(n)}\leq K}\q_{\mathcal{M}^{(n)}_i}(S_{i,T}^{(n)})\to  \QQ_i(S_i (T)\leq K) \quad \text{as}\ n\to\infty\ \text{for\ each}\ K>0.\]
\item $S_i (t)$ is an increasing function of $X_t$ for each $t$.
\end{enumerate}
We assume that the `local market prices of risk' in the models almost surely coincide:
\[\frac{\mu_1 (t) -r_1 (t, S_1 (t))}{\sigma_1 (t, S_1 (t))} = \frac{\mu_2 (t) - r_2 (t, S_2 (t))}{\sigma_2 (t, S_2 (t))},\quad 
0\leq t \leq T .\]
If the risk-neutral density $\q_2$ on $S_2 (T)$ is applied together with the physical laws $\phi_i$ of $S_i (T)$ in distribution matching method with the corresponding transform $\mathcal{K}$ (see \eqref{eq: value}), then the risk-neutral density 
$\q_1$ of $S_1$ can be represented as follows:
\begin{equation}\label{eq: transf}
\q_1 (x) = \mathcal{K}' (x)  \q_{2}(\mathcal{K}(x))
\end{equation}
and in particular the correct risk-neutral prices of derivatives are recovered:
\[\E_{\Q_1}(f(S_1 (T))) =\int f(x)\ \q_2 (\mathcal{K}(x))\ \frac{\phi_1 (x)}{\phi_2 (\mathcal{K}(x))}\d x .\]
\end{theorem}
Since the short rates may differ in the models, the AD securities may be differently scaled in these models and this is why we have a formula for risk-neutral prices, instead of state prices.

\section{Discussion}

We introduced a novel 'Distribution Matching' asset pricing technique which provides a natural correction to derivatives prices with respect to a benchmark model. The corrected RND (estimate) can be neatly expressed:
\[\widehat{\q}^{\mathrm{DM}}_1 (x) = \frac{\phi_{1}(x)}{\phi_{2}(\K(x))} \q_2 (\K(x)) . \]
This is operational in the sense that the densities $\phi_{1}, \phi_{2}, \q_2$ can be estimated or assumed from a given model
and $\K$ can be easily solved numerically.

In distribution matching one constructs essentially a European style derivative on a liquid proxy security. According to the static hedging principles this derivative will be correctly priced, per se, in the given framework. Of course, the value of a derivative need not be close in general to the value of an asset with the same future price distribution. However, if the derivative payoff and the asset price are additionally highly correlated, then there are various avenues of financial arguments suggesting that the values should be close as well. In fact, this principle seems to have been applied in the finance literature abundantly in this connection, although, usually somewhat implicitly. For some related works, apart from the Grahm-Charlier approach and references above, see Jackwerth and Rubinstein (1996) and Madan, Carr and Chang (1998).

Athough intended primarily as an approximate estimate, some example frameworks were given where the technique performs exactly correctly. The technique induces a RND estimate which is interesting on its own right and serves as a basis for further analysis of derivatives prices. Namely, using the distribution matching technique 
\emph{in reverse} with estimated state prices as an input one obtains 'Implied Physical Distributions', generalizing 
implied volatility. Recall in this connection the well-known Recovery Theorem of Ross (2015). 

In case the asset to be priced is essentially a derivative on the market index in a representative agent model, then 
the Lucas (1978) first order condition can in principle be applied in pricing the asset. However, this requires the knowledge of the exact form of the representative utility function. In practice there are some problematic issues with this approach. For instance, the Equity Premium Puzzle by Mehra and Prescott (1985) involves risk aversion constants in CRRA utility functions which are higher than anticipated by behavioral empirical studies. Even worse, Bakshi et al. (2010) observe that some empirically observed pricing kernels, which should coincide with marginal utility in respective states at the equilibrium, implies that the corresponding utility function should strongly fail to be concave. Therefore flexible calibration techniques are required instead.

One is tempted simply using the Lucas first order condition anyway, outside the equilibrium framework, with a model-$2$ pricing kernel $\frac{\d\Q_2}{\d\P_2}$, say of a BSM model. Namely, in some frameworks under equilibrium prices one may equate a marginal utility $u' (S_2 (T)) = \frac{\d\Q_2}{\d\P_2}(S_2 (T))$ up to scaling, see A\"{\i}t-Sahalia and Lo (2003), Breeden and Litzenberger (1978). Therefore it seems promising to analyze the expectation
\[\E_{\P_1} (u' (S_2 (T)) f(S_1 (T))) \stackrel{?}{=}  \E_{\P_1} \left(  \frac{\d\Q_2}{\d\P_2}(S_2 (T))    f(S_1 (T))\right) .\]
However, this does not result in the correct BSM model-$1$ risk-neutral value of the option, $\E_{\Q_1} (f(S_1 (T)))$, even if the states formally coincide, 
$S_1 (T) =S_2 (T)$. See Example \ref{example}.
The point is that 
$\frac{\d\Q_2}{\d\P_2} \d\P_1 \neq \d\Q_1$ and changing the market's physical distribution affects the equilibrium as well, 
even if risk preferences remain the same. Instead, the risk-neutral measures must be reconciled more carefully. 
Theorem provides an interpretation for the state space transformation $\mathcal{K}$ investigated here. The transform may reflect a Girsanov style change of dynamics between risk-neutral processes.  
 
We discussed distribution matching both with SPDs and RNDs. The former is mildy simpler. 
The latter is more flexible in the sense that the compared pricing models may then have different short rates or discounting terms. This may be a useful feature if the short rates in the models must be adjusted such that the resulting market prices of risk coincide. 


Here we were working essentially with one market index only. Relying on an analogy where the version of the technique investigated here corresponds to CAPM, one may ask about a multifactor case, in the spirit of the APT. Thus, future work may include extending the pricing technique to a multidimensional state price density, where the dimensions may 
correspond to macroeconomical indicators. Other future work may include more generally differential asset pricing, where there is a benchmark state price density available. These may include corporate bonds with respect to stocks and government bonds, index options with respect to other highly correlated index options (e.g. SP100 vs. SP500), and commodity derivatives with respect to related commodity derivatives.

\section*{Appendices}

\begin{appendices}

\section{Proofs}\label{sect: proofs}
We will give sketches of proofs retaining the notations and assumptions appearing in the statements.

\begin{proof}[Proof of Proposition \ref{prop: properties}]
For each $n\in\N$ let $\{x_{k}^{(n)}:\ 1\leq k\leq n\}$ and $\{y_{k}^{(n)}:\ 1\leq k\leq n\}$ be increasing partitions 
of the supports of $\phi_1$ and $\phi_2$ (possibly with $y_{n}^{(n)}=\infty$) such that 
\[\int_{x_{k}^{(n)}}^{x_{k+1}^{(n)}}\phi_1 (x)\d x=\int_{y_{k}^{(n)}}^{y_{k+1}^{(n)}}\phi_2 (y)\d y=1/n\] 
for all $1\leq k <n$.

It follows from the properties of $\phi_1$ that $\sup_{k} (x_{k+1}^{(n)} - x_{k}^{(n)}) \to 0$ as $n\to \infty$.
By compactness considerations we observe that 
\[\min\{\phi_{2}(y):\ \mathcal{K}(x_{2}^{(n)})=y_{2}^{(n)}\leq y\leq y_{n-1}^{(n)}=\mathcal{K}(x_{n-1}^{(n)})\}\] 
exists and is non-zero. Therefore 
the function 
\[x\mapsto \frac{\phi_1 (x)}{\phi_2 (\K(x))},\]
defined on $[x_{2}^{(n)}, x_{n-1}^{(n)}]$ is uniformly continuous. This means that
\begin{equation}\label{eq: lim-sup}
\lim_{n\to\infty}\sup_{\max(x_{k}^{(n)},x_{2}^{(i)})\leq x \leq \min(x_{k+1}^{(n)},x_{i-1}^{(i)})} 
\left| \mathcal{K}'(x) - \frac{x_{k+1}^{(n)}- x_{k}^{(n)}}{y_{k+1}^{(n)}-y_{k}^{(n)}}\right|=0
\end{equation}
for $i \in \N$. 

Note that by the integrability of $q$ we have that
\[\frac{1}{y_{k+1}^{(n)}- y_{k}^{(n)}} \int_{x_{k}^{(n)}}^{x_{k+1}^{(n)}} (x_{k+1}^{(n)}-x_{k}^{(n)}) q(\K(x))\ \d x\to 0\]
as $n\to \infty$ for every $k$ (including cases $k=1,n-1$). 

Since $q$ is integrable and $\mathrm{supp}(\phi_1 )$ bounded, we have by \eqref{eq: lim-sup} that 

\[\lim_{n\to \infty} \sum_{k=2}^{n-2}\frac{x_{k+1}^{(n)}+x_{k}^{(n)}}{2}\left|\int_{y_{k}^{(n)}}^{y_{k+1}^{(n)}}q(K)\ \d K 
-\int_{x_{k}^{(n)}}^{x_{k+1}^{(n)}}\frac{x_{k+1}^{(n)}-x_{k}^{(n)}}{y_{k+1}^{(n)}-y_{k}^{(n)}}\ q(\K(x))\ \d x\right|=0\]
and
\begin{multline*}
\lim_{n\to \infty} \sum_{k=2}^{n-2}\frac{x_{k+1}^{(n)}+x_{k}^{(n)}}{2}\int_{x_{k}^{(n)}}^{x_{k+1}^{(n)}}\frac{x_{k+1}^{(n)}-x_{k}^{(n)}}{y_{k+1}^{(n)}-y_{k}^{(n)}}\ q(\K(x))\ \d x\\
=\int x \frac{\phi_{1}(x)}{\phi_{2}(K(x))} q(\K(x))\ \d x.
\end{multline*}
from which the claim follows.
\end{proof}

\begin{proof}[Proof of Proposition \ref{prop: idemp}]
Observe that $\mathcal{K}$ is necessarily an identical mapping and $\frac{\phi_{1}(x)}{\phi_{2}(\K(x))}=1$ for $\P_1$-a.e. $x$.
By using the definition of (the absence of) arbitrage we obtain that 
\[\Pi_{\rho}=\int x\ q_2(\mathcal{K}(x))\ \d x=\int x\ q_2(x)\ \d x=S_2 (t).\]
\end{proof}

\begin{proof}[Proof of Proposition \ref{prop: monotonicity}]
We will follow the finite approximation of the portfolios as follows. Let $(x_{k})_{k}$, $(y_{k})_{k}$
and $(z_{k})_{k}$ be increasing sequences of $\R$ such that $F_{a}(x_k)= F_{b}(y_k)$ 
for each $k$. Then $x_k\leq y_k$ by the assumption. Thus
\begin{equation}
\sum_{k\in \mathbb{Z}}\frac{x_{k+1}+x_{k}}{2}\int_{z_{k}}^{z_{k+1}} q(K)\ \d K \leq  \sum_{k\in\mathbb{Z}} \frac{y_{k+1}+y_{k}}{2}\int_{z_{k}}^{z_{k+1}} q(K)\ \d K.
\end{equation}
\end{proof}

\begin{proof}[Proof of Proposition \ref{prop: cont}]
Indeed, let $\sup_{K}q(K)=C$. We will apply  \eqref{eq: value}. We will compare the values
\[\int x\ \widehat{q}^{\DM}_1 (x) \d x\quad \text{and}\quad \int x\ \widehat{q}^{\DM}_{1, (n)}(x) \d x.\]
Since we are dealing with pricing measures we may restrict to analysing intervals of the form $[\varepsilon, \infty)$ where 
$\varepsilon>0$. Fix $N>\varepsilon$. By using the unimodality of $\phi_2$ and the selection $C$ it follows that there is an upper bound $C_1 >0$ for $\frac{q_2 (y)}{\phi_2 (y)}$ on the interval $[\varepsilon, N)$ such that 
\begin{multline*}
\lim_{n\to\infty} \left|\int_{\varepsilon}^N x\ \widehat{q}^{\DM}_1 \d x - \int_{\varepsilon}^N x\ \widehat{q}^{\DM}_{1, (n)} \d x\right|\\
=\lim_{n\to\infty}  \left|\int_{\varepsilon}^N x\ \frac{q_2 (\mathcal{K}_1 (x))}{\phi_2 (\mathcal{K}_1 (x))}\phi_{1}\d x 
- \int_{\varepsilon}^N x\ \frac{q_2 (\mathcal{K}_{1, (n)} (x))}{\phi_2 (\mathcal{K}_{1,(n)} (x))}\phi_{1,(n)}\d x \right|\\
=\lim_{n\to\infty} \left|\int_{\varepsilon}^N x\ \frac{q_2 (\mathcal{K}_1 (x))}{\phi_2 (\mathcal{K}_1 (x))} \d F_{1} (x) - \int_{\varepsilon}^N x\ \frac{q_2 (\mathcal{K}_1 (x))}{\phi_2 (\mathcal{K}_1 (x))} \d F_{1,(n)}(x) \right|\\
\leq \lim_{n\to\infty} C_1  \left|\int_{\varepsilon}^N x\ \d F_{1} (x)- \int_{\varepsilon}^N x\ \d F_{1, (n)} (x)\right|\\
=\lim_{n\to\infty} C_1 |\E_{\P_1} (1_{\varepsilon \leq S_1 \leq N} S_1 - 1_{\varepsilon \leq S_{1,(n)} \leq N} S_{1, (n)})|=0,
\end{multline*} 
since $S_{1,(n)} (T) \to S_1$ in the $\P_1$-mean. The argument is finished by using the fact that 
$\frac{q_2 (y)}{\phi_2 (y)}\to 0$ as $y\to\infty$.
\end{proof}

\begin{proof}[Proof of Proposition \ref{prop: contclaim}]
It follows from the assumptions that we may consider $\mathcal{K} \colon f_{S_1} (S_2 (T)) \mapsto S_2 (T)$. 
Then 
\begin{eqnarray*}
V_{S_1} &=&\int x \frac{\phi_1 (x)}{\phi_{2}(\K(x))} q_2 (\K(x))\d x =\int x\ \K'(x) \frac{\d \Q_2}{\d K} (\K(x))\d x\\
&=&\int x \frac{\d \Q_2}{\d x} (\K(x))\d x =\int f_{S_1} (S_2 (T)) \d \Q_2 (S_2 (T)).
\end{eqnarray*}
This is the risk-neutral price of the contingent claim with payoff $f_{S_1}$.
\end{proof}

\begin{proof}[Proof of Theorem \ref{thm: coincide_MPR}]
According to the previous lemma we see that the states $S_{1}^{(n)} (T)=a_1 , a_2 , \ldots ,a_k$ and $S_{2}^{(n)} (T)=b_1 , b_2 , \ldots , b_k$, written in an increasing order and with $k \leq 2^n$, have equal probabilities: 
\[\mathbb{P}( S_{1}^{(n)} (T)=a_j)= \mathbb{P}( S_{2}^{(n)} (T)=b_j),\quad 1\leq j \leq k.\]
Therefore the assumption regarding the convergence of the discretized versions of 
$S_i$ yield with an easy approximation argument that 
\[\mathbb{P}(S_1 (T) \leq x) = \mathbb{P}(S_2 (T) \leq \mathcal{K}(x))\]
for each $x>0$ where $\mathcal{K}$ is the $S_1$-state-to-$S_2$-state binding map appearing in the definition of $\Pi_{\rho}$.

What remains to be verified is that the following holds:
\begin{equation}\label{eq: char}
\frac{\d\QQ_1 (x)}{\d\mathbb{P}(S_1 (T)=x)}=\frac{\d\QQ_2 (\mathcal{K}(x))}{\d\mathbb{P}(S_2 (T)=\mathcal{K}(x))}
\end{equation}
for all $x=S_1 (T)>0$. Note that then
\[\E_{\Q_1} (f(S_1 (T))) = \int_{0}^\infty f(x) \q_2 (\mathcal{K}(x)) \frac{\d\mathbb{P}(S_1 (T)=x) }{\d\mathbb{P}(S_2 (T)=\mathcal{K}(x))} \d x.\]

Recall the simple well-known  equalities of discounted risk-neutral probabilities in a single-step model 
(see e.g. F\"ollmer and Schied (2011)): 
\[q(u)=e^{-r\vartriangle t}\ \frac{e^{r\vartriangle t}-d}{u-d},\quad q(d)=e^{-r\vartriangle t}\ \frac{u-e^{r\vartriangle t}}{u-d}.\]
Consider the discounted price processes of the securities $S_i$
\[\frac{d \widetilde{S}_i}{\widetilde{S}_i} = \sigma_i (S_i (t)) dX_t +  (\mu_i (t) - r_i (t))dt \]
and their binomial counterparts. Denote $t_k = k\! \vartriangle\!\! t$. 
Then in the binomial discrete models for $i=1,2$ the single-step subtrees have the same risk-neutral 
probabilities. Indeed, in the discounted world the risk-neutral terms corresponding to 
$u-r=\widetilde{u}-0$, $r-d=0-\widetilde{d}$ and $u -d$ become 
\[(\mu_i (t_k ) - r_i (t_k , S^{(n)}_{i, t_k}) + \sigma_i (S^{(n)}_{i, t_k})\Theta_{t_k}^{(n)\ \uparrow} ) ,\]
\[0- (\mu_i (t_k) - r_i (t_k , S^{(n)}_{i, t_k}) + \sigma_i (S^{(n)}_{i, t_k} ) \Theta_{t_k}^{(n)\ \downarrow}) \]
and
\[(\mu_i (t_k ) + \sigma_i (S^{(n)}_{i, t_k})\Theta_{t_k}^{(n)\ \uparrow} ) -(\mu_i (t_k) + \sigma_i (S^{(n)}_{i, t_k})\Theta_{t_k}^{(n)\ \downarrow} ) ,\]
respectively. We use arrows to indicate the change of a state. Thus
\[\QQ_{n,i} (S^{(n)}_{i, t_k} \uparrow\  |\ \mathcal{F}_{t_k})  
= \frac{- (\mu_i (t_k ) - r_i (t_k , S^{(n)}_{i, t_k}) + \sigma_i (S^{(n)}_{i, t_k})\Theta_{t_k}^{(n)\ \downarrow} )}{\sigma_i (S^{(n)}_{i, t_k})(\Theta_{t_k}^{(n)\ \uparrow} - \Theta_{t_k}^{(n)\ \downarrow})} ,\]
\[\QQ_{n,i} (S^{(n)}_{i, t_k} \downarrow\  |\ \mathcal{F}_{t_k}) 
= \frac{\mu_i (t_k ) - r_i (t_k , S^{(n)}_{i, t_k}) + \sigma_i (S^{(n)}_{i, t_k}) \Theta_{t_k}^{(n)\ \uparrow} }{ \sigma_i (S^{(n)}_{i, t_k})(\Theta_{t_k}^{(n)\ \uparrow} - \Theta_{t_k}^{(n)\ \downarrow}) } .\]
 
Consequently, under the assumption on coinciding local market prices of risk we obtain that
\[\QQ_{n,1} (S^{(n)}_{1, t_k} \uparrow \ |\  \mathcal{F}_{t_k}) = \QQ_{n,2} (S^{(n)}_{2, t_k} \uparrow \ |\ \mathcal{F}_{t_k}),\]
\[\QQ_{n,1} (S^{(n)}_{1, t_k} \downarrow \ |\ \mathcal{F}_{t_k}) = \QQ_{n,2} (S^{(n)}_{2, t_k} \downarrow \ |\ \mathcal{F}_{t_k}) .\]
We conclude that the binomial trees corresponding to the assets $i=1,2$ are isomorphic to that of 
$X^{(n)}$, since the orders are preserved, and, moreover, both the physical probabilities and state prices of the corresponding nodes coincide. 
This means that 
\[\frac{\mathbb{P}(S^{(n)}_1 (T) = a_j)}{\QQ_{n,1}(S^{(n)}_1 (T) = a_j)}  = \frac{\mathbb{P}(S^{(n)}_2 (T) = b_j)}{\QQ_{n,2}(S^{(n)}_2 (T) = b_j)}\]
for the mutually corresponding terminal node state values $a_j$ and $b_j$ 
(in the trees of the respective models) for each $j$. A straight-forward approximation argument then yields the 
claim that \eqref{eq: char} holds and this finishes the proof.
\end{proof}

\end{appendices}

\end{document}